\documentstyle[12pt]{article}
\newlength{\extraspace}
\newlength{\extraspaces}
\newcounter{savefootnote}
\begingroup
\endgroup

\newcommand{\be}{\begin{equation}
\addtolength{\abovedisplayskip}{\extraspaces}
\addtolength{\belowdisplayskip}{\extraspaces}
\addtolength{\abovedisplayshortskip}{\extraspace}
\addtolength{\belowdisplayshortskip}{\extraspace}}
\newcommand{\ee}{\end{equation}}

\newcommand{\ba}{\begin{eqnarray}
\addtolength{\abovedisplayskip}{\extraspaces}
\addtolength{\belowdisplayskip}{\extraspaces}
\addtolength{\abovedisplayshortskip}{\extraspace}
\addtolength{\belowdisplayshortskip}{\extraspace}}
\newcommand{\ea}{\end{eqnarray}}

\newcommand{\nonu}{\nonumber \\[.5mm]}
\newcommand{\A}{&\!\!\!}

\def\thesection {\S {\arabic{section}}}
\newcommand{\newsection}[1]{
\vspace{7mm} \pagebreak[3] \addtocounter{section}{1}
\setcounter{subsection}{0} 
\begin{center}
{\large {\bf \thesection. #1}}
\end{center}
\nopagebreak
\medskip
\nopagebreak \hspace{3mm}}

\begin{document}

\begin{center}
{\bf   A spherically-symmetric charged-dS solution in f(T) gravity theories}\footnote{ PACS numbers: 04.50.Kd, 04.70.Bw, 04.20. Jb.\\
\hspace*{.5cm}
 Keywords: f(T) theories of gravity, special exact spherically-symmetric charged-dS solution.}
\end{center}
\begin{center}
{\bf Gamal G.L. Nashed}
\end{center}

\bigskip

\centerline{\it Centre  for Theoretical Physics, The British
University in  Egypt} \centerline{\it Sherouk City 11837, P.O. Box
43, Egypt \footnote{ Mathematics Department, Faculty of Science, Ain
Shams University, Cairo, 11566, Egypt. \\
\hspace*{.2cm} Egyptian Relativity Group (ERG) URL:
http://www.erg.eg.net.}}

\bigskip
\centerline{ e-mail:nashed@bue.edu.eg}

\hspace{2cm} \hspace{2cm}
\\
\\
\\
\\
\\

A  tetrad field with  spherical symmetry  is applied to the
charged field equations of $f(T)$ gravity theory. A special
spherically-symmetric charged-dS solution is obtained. The scalar
torsion of this solution is  a vanishing quantity. The spacetime of
the derived solution is rewritten as a multiplication  of three
matrices: The first matrix is a special case of Euler$'$s angle
``so(3)", the second matrix  represents a boost transformation, while the third
matrix is the square root of the spherically-symmetric charged-dS
metric. It is shown that the boost matrix is  important  because
it plays an essential role in adjusting the spacetime to become a solution
for $f(T)$ theory.
\begin{center}
\newsection{\bf Introduction}
\end{center}

Extensions of General Relativity (GR) aim to demonstrate the
late-time acceleration of the universe and Dark Energy (DE). Recent
 observations suggest the presence of fluids as well as
the known pressureless matter terms which permit our universe to
exhibit a  late-time acceleration \cite{Ps,Rg,Rr}. At present, the physical existence of such  fluids is  unknown
\cite{CS}. An intensive discussion has
emerged  to understand  any form of fluids that can cause the
observed  cosmic acceleration \cite{CST6}. Although its physical
nature is not known,  it displays an Equation of State (EoS) with negative pressure
\cite{Pt,Tm,Ws8}. This EoS  supplies an antigravity impact that
balances  the pull of gravity acting on standard matter
\cite{LQ,Ts, W12}. This hypothetical fluid, usually named as DE, is representing
 more than $70\%$ of  all the universe energy budget
\cite{Pd6}.

Recently, many efforts have been devoted to the amended theories of
gravity that unify the method of early-time inflation and late-time
acceleration of the universe. Such theories possibly supply another
framework that features our understanding of DE. The gravitational action in
such theories should apply at the low-energy limit of fundamental
quantum gravity. These gravitational theories differ from GR
  (for example $f(R)$), where $R$ is the Ricci scalar in the
Einstein-Hilbert action \cite{SF,FT}. In GR and its amendments the metric $g_{\mu \nu}$ and the quantities derived from it
characterize the gravitational field.

The basic  philosophy of the amended theories of gravity is that GR must
be viewed as a special case of a  wider theory derived from basic
principles \cite{CLF}-\cite{CF8}. The underlying idea is that  the
 standard Einstein-Hilbert action is changed by adding
degrees of freedom. These include extra curvature-invariant
corrections to scalar fields and Lorentz-violating terms.

Most recently, $f(T)$ gravitational theories  were constructed  as an alternative to the cosmological
constant, to give an explanation to  the accelerated expansion of the universe \cite{BF9}. By analogy  with
the $f(R)$ gravitational  theories, modified teleparallel equivalent of general relativity  theories differ  from GR by a function $f(T)$ in the
Lagrangian, where $T$  is called the torsion scalar. This set of theories has gained a lot
of attention in the last few years \cite{CCT}-\cite{BCN}.  The
 $f(T)$  gravitational theories are particularly  attractive since their field
equations are of second order, rather than fourth order as in $f(R)$. The aim of this study is to find an analytic,
spherically-symmetric charged-dS solution  within the framework of
$f(T)$ gravity theory.

 $f(T)$ gravitational theories represent a category of models, which takes into account the influences related
 to the torsion scalar, $T$.  The feature of $f(T)$ gravity is that the curvature tensor
  identically vanishes using the Weitzenb\"ock connection.   $f(T)$ model can be viewed as expansion of  ``teleparallel gravity" \cite{CCT}.
  Therefore, one can amend the Lagrangian density by adding a torsion function  $f(T)$.

 In \S 2, a  survey of the $f(T)$
gravitational theory is supplied.  A non-diagonal,
spherically-symmetric tetrad field  is provided.

Application of charged equations of motion, of $f(T)$ gravitational
theory, to the non-diagonal tetrad field is given  in \S 3. The
special analytic, non-vacuum, spherically-symmetric charged-dS
solution is derived with two constants of integration.

 In \S 4, the internal properties of the derived solution is
 demonstrated.

 In the last section, a discussion of the main results is given.
\newpage
\newsection{Brief review of f(T)}
In a spacetime having  absolute parallelism, the tetrad field ${h_i}^\mu$ \cite{Wr} defines the nonsymmetric affine
connection: \be {\Gamma^\mu}_{\nu \lambda} \stackrel{\rm def.}{=} {h_i}^\mu
{h^i}_{\nu, \lambda}, \ee where ${h^i}_{\nu, \lambda}=\partial_\lambda {h^i}_{\nu}$\footnote{We use the Greek indices ${\it \mu, \nu, \cdots }$ for
local holonomic spacetime coordinates and the Latin indices
$i$, $j$, $\cdots$ to label (co)frame components.}.

The
curvature tensor defined by ${\Gamma^\mu}_{\nu  \lambda}$, given by Eq. (1),
is identically vanishing. The metric tensor $g_{\mu \nu}$
 is defined by:
 \be g_{\mu \nu} \stackrel{\rm def.}{=}  \eta_{ i j} {h^i}_\mu {h^j}_\nu, \ee
with $\eta_{\mu \nu}=(+1,-1,-1,-1)$ denoting the metric of Minkowski spacetime.
 The torsion components and the contortion are given, respectively, as: \ba
{T^\mu}_{\nu \lambda}  \A \stackrel {\rm def.}{=} \A
{\Gamma^\mu}_{\lambda \nu}-{\Gamma^\mu}_{\nu \lambda} ={h_i}^\mu
\left(\partial_\nu{h^i}_\lambda-\partial_\lambda{h^i}_\nu\right),\nonu
{K^{\mu \nu }}_\lambda  \A \stackrel {\rm def.}{=} \A
-\frac{1}{2}\left({T^{\mu \nu }}_\lambda-{T^{\nu
\mu}}_\lambda-{T_\lambda}^{\mu  \nu}\right), \ea  where the contortion
equals the difference between Weitzenb\"ock  and
Levi-Civita connection, i.e., ${K^{\mu}}_{\nu \lambda}= {\Gamma^\mu}_{\nu
\lambda }-\left \{_{\nu  \lambda}^\mu\right\}$.

The tensor ${S_\mu}^{\nu \lambda}$ is defined as: \be {S_\mu}^{\nu \lambda} \stackrel
{\rm def.}{=} \frac{1}{2}\left({K^{\nu \lambda}}_\mu+\delta^\nu_\mu{T^{\beta
\lambda}}_\beta-\delta^\lambda_\mu{T^{\beta \nu}}_\beta\right). \ee The torsion scalar is
defined as: \be T \stackrel {\rm def.}{=} {T^\mu }_{\nu \lambda} {S_\mu}^{\nu \lambda}.
\ee Similar to the $f(R)$ theory, one can define the action of $f(T
)$ theory as \ba \A \A {\cal L}({h^a}_\mu)=\int d^4x
h\left[\frac{1}{16\pi}(f(T)-2\Lambda)+{\cal L}_{ em}\right],  \nonu
\A \A \textrm
{where} \quad h=\sqrt{-g}=\det\left({h^a}_\mu\right),\nonu
\A \A   {\cal L}_{ em} \quad \textrm {is \quad  the
\quad  Lagrangian \quad  of \quad  electromagnetic \quad  field}\nonu
\A \A \textrm {and} \quad  \Lambda \quad \textrm {is \quad  the
\quad  cosmological  \quad  constant.}\ea

We are going to use the units in which $G = c = 1$ and ${\cal L}_{
em}=-\frac{1}{2}{ F}\wedge ^{\star}{F}=-\displaystyle{1 \over 4}
g^{\mu \rho} g^{\nu \sigma} F_{\mu \nu} F_{\rho \sigma}$ is the
Maxwell Lagrangian, while\footnote{We will denote the
symmetric part by (\ ), for example, $A_{(\mu \nu)}=\frac{1}{2}(A_{\mu
\nu}+A_{\nu \mu})$ and the antisymmetric part by the square
bracket [ \ ],  $A_{[\mu \nu]}=\frac{1}{2}(A_{\mu \nu}-A_{\nu \mu})$.} $F =
dA=A_{[\mu,\nu]}[dx^\mu \wedge dx^\nu] $,
with $A=A_{\mu}dx^\mu$, is the electromagnetic potential 1-form
\cite{CGSV3}.  Similar to the $f(R)$ theory, one can define the
action of $f(T )$ theory as   a function of the fields ${h^a}_\mu$
and, by putting  the variation of the function with respect to the tetrad
field ${h^a}_\mu$ to be vanishing, we can obtain the following
equations of motion: \be {S_\mu}^{\rho \nu} T_{,\rho} \
f(T)_{TT}+\left[h^{-1}{h^a}_\mu\partial_\rho\left(h{h_a}^\alpha
{S_\alpha}^{\rho \nu}\right)-{T^\alpha}_{\lambda \mu}{S_\alpha}^{\nu
\lambda}\right]f(T)_T-\frac{{\delta_\mu}^\nu(f(T)-2 \Lambda)}{4} =-4\pi
{{{\cal T}^{{}^{{}^{^{}{\!\!\!\!\scriptstyle{em}}}}}}}^\nu_\mu,\ee
\be
\partial_\nu \left( \sqrt{-g} F^{\mu \nu} \right)=0, \ee
 where
\[T_{,\rho}=\frac{\partial T}{\partial x^\rho}, \qquad \qquad
f(T)_T=\frac{\partial f(T)}{\partial T}, \qquad \qquad
f(T)_{TT}=\frac{\partial^2 f(T)}{\partial T^2}, \]  and
 \[ {{{\cal
T}^{{}^{{}^{^{}{\!\!\!\!\scriptstyle{em}}}}}}}^{\mu \nu}=g_{\rho
\sigma}F^{\mu \rho}F^{\nu \sigma}-\displaystyle{1 \over 4} g^{\mu
\nu} g^{\lambda \rho} g^{\epsilon \sigma} F_{\lambda \epsilon}
F_{\rho \sigma}, \] is the energy momentum tensor for
electromagnetic field.
\newsection{Spherically symmetric solution in  f(T) gravity theory}
Assuming that the manifold possesses a stationary and
 spherical symmetry, the tetrad field has the form:
 \be
\left( {h^i}_\mu \right)=  \left( \matrix{ { A}_1(r) &
{ A}_2(r) & { A}_3(r) & {A}_4(r)
 \vspace{3mm} \cr  { B}_1(r) \sin\theta
\cos\phi  & {B}_2(r) \sin\theta \cos\phi &
{ B}_3(r) \cos\theta \cos\phi
 & { B}_4(r) \sin\phi  \sin\theta \vspace{3mm} \cr
 { C}_1(r) \sin\theta \sin\phi  & { C}_2(r)  \sin\theta \sin\phi & { C}_3(r)
   \cos\theta \sin\phi & { C}_4(r)  \cos\phi \sin\theta \vspace{3mm} \cr
 { D}_1(r) \cos\theta&   { D}_2(r) \cos\theta & { D}_3(r)\sin\theta  &
  {D}_4(r) \cos\theta  \cr }
\right)\; , \ee
 where ${ A}_i(r)$, ${ B}_i(r)$, ${ C}_i(r)$ and  ${ D}_i(r)$, $i=1\cdots
 4$, are sixteen  unknown functions of the radial coordinate, $r$.
 The metric of spherically-symmetric  has the
 form \cite{Ort}:
 \be ds^2=A(r) dt^2-\frac{1}{A(r)}dr^2-r^2(d\theta^2+\sin^2\theta d\phi^2).\ee
 To create metric (10) from tetrad field (9) we must have \be { A}_3={ A}_4= {
 D}_4=0, \qquad  { B}_3= -{ B}_4={ C}_3=-{ D}_3={ C}_4=r\; , \qquad { B}_2={ C}_2= {
 D}_2\; , \qquad { B}_1={ C}_1= {  D}_1. \ee  Constraints (11) result from the use of spherical symmetry.
 Using equation (11) in tetrad field (9)  we get:
 \be
\left( {h^i}_\mu \right)_1= \left( \matrix{A_1(r) & A_2(r)& 0 &
0\vspace{3mm} \cr B_1(r)\sin\theta \cos\phi &B_2(r)\sin\theta
\cos\phi&r\cos\theta \cos\phi & -r\sin\theta \sin\phi \vspace{3mm}
\cr B_1(r)\sin\theta \sin\phi&B_2(r)\sin\theta \sin\phi&r\cos\theta
\sin\phi & r\sin\theta \cos\phi \vspace{3mm} \cr B_1(r)\cos\theta &
B_2(r)\cos\theta&-r\sin\theta & 0 \cr } \right).\ee The metric of tetrad field (12) takes the form
\be ds^2=\eta_{i j}\left( {h^i}_\mu \right)_1 \left( {h^j}_\nu \right)_1dx^\mu dx^\nu=\left[{A_1}^2(r)-{B_1}^2(r)\right] dt^2-\left[{A_2}^2(r)-{B_2}^2(r)\right] dr^2-r^2(d\theta^2+\sin^2\theta d\phi^2).\ee
 Using Eq. (7) and
  tetrad field (12) one can obtain $h = \det ({h^a}_\mu) = r^2\sin \theta
  (A_1B_2-B_1A_2):$
With the use of equations (3) and (4), we obtain the torsion scalar
and its derivatives in terms of $r$ as:\footnote{For brevity, we  write
${ A}_i(r)\equiv { A}_i$, \  ${ B}_i(r)\equiv { B}_i$, \
   ${C}_i(r)\equiv  { C}_i$ \ and  \  ${ D}_i(r)\equiv { D}_i$.} \ba \A \A
T(r)=-\frac{4\left(rA'_1[(B_2-1)A_1-B_1A_2]+rB_1B'_1-\frac{1}{2}[(B_2-1)A_1-B_1(A_2-1)][(B_2-1)A_1-B_1(A_2+1)]\right)}{r^2(A_1B_2-B_1A_2)^2},
\nonu
\A \A \nonu
\A \A  \textrm{where} \qquad A'_1=\frac{\partial A_1(r)}{\partial r},
\qquad A'_2=\frac{\partial A_2(r)}{\partial r}, \qquad B'_1=\frac{\partial
B_1(r)}{\partial r}, \qquad \textrm{and} \qquad B'_2=\frac{\partial
B_2(r)}{\partial r},\nonu
\A \A \nonu
\A \A T'(r)\equiv \frac{\partial T(r)}{\partial r}=-\frac{4}{r^3(A_1B_2-B_1A_2)^3}\Biggl(r^2(A_1B_2-B_1A_2)[\{(B_2-1)A_1-B_1A_2\}A''+B_1B''_1]\nonu
\A\A-r^2\{(B_2-1)B_2A_1-B_1A_2(B_2+1)\}A'^2+4rA'_1\Biggl[rB'_1\{A_2(B_2-2)A_1-B_1(A_2^2+2B_2)\}\nonu
\A \A-r\{B_1A'_2-A_1B'_2\}[(B_2-2)A_1-B_1A_2]-B_2\{(B_2-1)A_1^2-A_1B_1A_2+B_1^2\}\Biggr]+r^2(A_1B_2+B_1A_2)B'^2_1\nonu
\A \A -2rA'_2\Biggl[rA_1B_1B'_2-rB_1^2A'_2-\frac{A_2}{2}\{(B_2-1)A_1^2-A_1A_2B_1+B_1^2\}\Biggr]-r[(B_2-1)A_1^2-A_1A_2B_1+B_1^2]\nonu
\A \A (A_1B'_2-B_1A'_2)+(A_1B_2-B_1A_2)[(B_2-1)A_1-(A_2-1)B_1][(B_2-1)A_1-(A_2+1)B_1]\Biggr).\ea
 The equations of motion (7) can be rewritten in the form \ba \A \A
4\pi{\cal T}_0^0=-\frac{f_{TT}T'[(B_2-1)A_1^2-A_1A_2B_1+B_1^2]}{r(A_1B_2-B_1A_2)^2}
-\frac{f_{T}}{r^2(A_1B_2-B_1A_2)^3}\Biggl(rA'_1\{A_1(B_2-1)(B_2A_1-2B_1A_2)\}\nonu
\A \A +B_1^2(A_2^2-B_2)-rB_1'(A_1^2A_2+A_2B_1^2-2A_1B_1B_2)+r(A_1B'_2-B_1A'_2)(A_1^2-B_1^2)\nonu
\A \A +(A_1B_2-B_1A_2)[(B_2-1)A_1^2-A_1A_2B_1+B_1^2]\Biggr)
+\frac{f-2\Lambda }{4}, \ea \be 4\pi{\cal T}_0^1=-\frac{2B_1f_{T}
\sin^2\theta \{A_1^2+A_1A_2B_1-B_1^2-rB_1B'_1-B_2A_1^2+rA_1A'_1
\}}{r^2(A_1B_2-B_1A_2)^2},\ee
 \be 4\pi{\cal
T}_1^0=
\frac{4f_{TT}T'(B_1A_2^2-A_1(B_2-1)-B_1B_2)}{r(A_1B_2-B_1A_2)^2}-\frac{f_{T}[A_1A_2-B_1B_2](B_2A'_1-A_2B'_1+A_1B'_2-B_1A'_2)}{r(A_1B_2-B_1A_2)^3},\ee
\be 4\pi{\cal T}_1^1=-\frac{f_{T}
\{B_1^2-A_1A_2B_1+(B_2-1)A_1^2+2rB_1B'_1+r[(B_2-2)A_1-B_1A_2]A'_1
\}}{r^2(A_1B_2-B_1A_2)^2}+\frac{f-2\Lambda}{4},\ee \be 4\pi{\cal
T}_2^0=-\frac{2f_{T} \sin2\theta
\{B_1A_2^2-A_1A_2B_2-B_2B_1-rB_2B'_1+A_1A_2+rA_2A'_1
\}}{r(A_1B_2-B_1A_2)^2},\ee
 \ba \A \A
4\pi{\cal T}_2^2=4\pi{\cal
T}_3^3=\frac{f_{TT}T'(r[A_1A'_1-B_1B'_1]-[B_2-1]A_1^2+A_1A_2B_1-B_1^2)}{2r(A_1B_2-B_1A_2)^2}\nonu
\A \A+\frac{f_{T}}{2r^2(A_1B_2-B_1A_2)^3}\Biggl(r^2(A_1B_2-B_1A_2)[A_1A''_1-B_1B''_1]-r^2B_1A_2
A'^2_1-rA'_1\Biggl[rB'_1\{B_1B_2+A_1A_2\}\nonu
\A \A-rA_1\{A_1B'_2-B_1A'_2\}+A_1^2B_2(3-2B_2) -
4A_1A_2B_1(1-B_2)+(B_2-2A_2^2)B_1^2\Biggr]\nonu
\A \A+r^2A_1B_2B'^2_1-rB'_1\Biggl(rA_1B_1B'_2-r B_1^2A'_2 + A_1^2A_2-4A_1B_1B_2+3
A_2B_1^2\Biggr)+r(A_1B_2'-B_1A'_2)(B_1^2-A_1^2)\nonu
\A \A
-(A_1B_2-B_1A_2)[(B_2-1)A_1-(A_2-1)B_1][(B_2-1)A_1-(A_2+1)B_1]\Biggr)+\frac{f-2 \Lambda }{4}.\ea When $f(T)=T$, i.e., $f_{TT}=0$ and $f_{T}=1$, Eqs. (13)--(19)
will be identical with the teleparallel equivalent of GR.
From equations (15)--(20), it is clear that $A_1B_2\neq B_1A_2.$
Equations (14)--(20) cannot be easily solved. This is because of the
existence of the two terms $f_T$ and $f_{TT}$. If these two terms
are vanishing then equations (15)--(20) can be easily  solved.
Therefore, to find an exact solution of equations (15)--(20), we put
the following constraints:  \ba \A \A T'=0,\nonu
\A \A A_1^2+A_1A_2B_1-B_1^2-rB_1B'_1-B_2A_1^2+rA_1A'_1=0,\nonu
\A \A B_2A'_1-A_2B'_1+A_1B'_2-B_1A'_2=0,\nonu
\A \A B_1A_2^2-A_1A_2B_2-B_2B_1-rB_2B'_1+A_1A_2+rA_2A'_1=0.\ea
Constraints (21) are resulting from the fact that we put the
derivatives of the scalar torsion to be vanishing, to attain  the
 removed of $f_{TT}$ terms from  the field equations (7).  The
other constraints in Eq. (21) come from the coefficient of $f_T$ in
the off-diagonal components  of Eqs. (16), (17) and (19). The constraints in  Eq.~(21) are adequate to secure the existence of special analytic
solution to Eqs. (15)--(20), as  will be show below. Equations (21) are
four differential equations in four unknown functions $A_1$, $A_2$,
$B_1$ and $B_2$. The only solution that can satisfy equations (21)
is the following\footnote{These calculations were checked using
Maple ``15" software.}.
 \ba { A}_1 \A
=\A 1+\frac{3{c_2}^2-6c_1r-r^4\Lambda}{6r^2}\; ,
\qquad \qquad {A}_2=\displaystyle\frac{3{c_2}^2-6c_1r-r^4\Lambda }{2\Delta}\;
, \qquad  {B}_1=1-{ A}_1 \; , \nonu
  {B}_2\A=\A-\displaystyle\frac{6r^2+[3{c_2}^2-6c_1r-r^4\Lambda ]}{2\Delta}\; ,\
 \ea
 where $\Delta,$ is defined by:
  \be
 \Delta =-{3r^2+3{c_2}^2-6c_1r-r^4\Lambda}\; .\ \ee
     Here $c_1$ and $c_2$ are two
constants of integration.

The tetrad field (12), after  using equation (22), is
axially-symmetric.
 This means that it is invariant under the transformation:
\ba \A \A \bar{\phi}\rightarrow \phi+\delta \phi\; , \qquad \qquad
{\bar{h}^0}_{\ \mu} \rightarrow {{h}^0}_{\mu}\; , \qquad \qquad {\bar{h}^1}_{\ \mu}
\rightarrow {{h}^1}_{\mu}\cos \delta \phi
 -{{h}^2}_{ \mu}\sin \delta \phi\; , \nonu
 \A \A {\bar{h}^2}_{\ \mu} \rightarrow {{h}^1}_{ \mu}\sin \delta \phi
 +{{h}^2}_{ \mu}\cos \delta \phi\; , \qquad \qquad \qquad {\bar{h}^3}_{\ \mu} \rightarrow {{h}^3}_{ \mu}\; . \ea

Using equations (12) and (22) in equation (5) we get a vanishing value of the scalar torsion, $T$, because of constraints (21). Substituting
 (12) and (22) in Eq.~(7), we get an exact non-vacuum solution to the field equations of $f(T)$ given by equations (7) and (8),  provided that \ba
\A \A f(0)=0,
 \qquad f_T(0)=1, \qquad f_{TT}\neq 0, \qquad A_t(r)=-\frac{{c_2}}{r\sqrt{2\pi}}, \qquad
F_{tr}=\frac{{c_2}}{r^2\sqrt{2\pi}},\nonu
\A \A
{T_0}^0={T_1}^1=-{T_2}^2=-{T_3}^3=-\frac{{c_2}^2}{4r^4\pi }.\ea
In summary: Solution (22) is a special analytic solution to the field
equations of $f(T)$ whatever the form of $f(T)$ is provided that Eqs.
(25) are satisfied. Equation (25) reduces to
the vacuum case when the constant
$c_2=0$ \cite{RR11}.
 To understand the physical meaning of the constants of integration that appear in solution (22)
  we discuss the physics related to this solution.
  First, the metric associated with the tetrad field given by equation (12)
  (after using solution  (22) in Eq. (13)) has the form:
  \be
  ds^2= \alpha dt^2 -\frac{1}{\alpha}dr^2 -r^2 d\Omega^2, \quad
with \quad  d\Omega^2=d\theta^2+\sin^2\theta d\phi^2,\qquad
 \alpha=\left(1-\frac{2c_1}{r}+\frac{{c_2}^2}{r^2}-\frac{r^2\Lambda}{3}\right).
\ee This is similar to Reissner Nordstr\"om-dS spacetime provided that
$c_1=M$ and $c_2=q$. When $\Lambda\rightarrow 0$, Eq.~(26) reduces to  Reissner Nordstr\"om spacetime.
Further, when  $\Lambda\rightarrow 0$  and $c_2=0$ then  Eq.~(26)  reduces to Schwarzschild spacetime and,  finally, when $c_2=0$
then Eq.~(26)  gives Schwarzschild-dS spacetime. All these special cases can be verified from solution (22).

\newsection{Internal properties}
In this section, we  rewrite the tetrad field (12) using
equation (22) in terms of a special case of Euler$'$s angle ``so(3)"
and a boost transformation. For this purpose, we use the fact that
any tetrad field can be written as: \be \left( {h^i}_\mu
\right)\stackrel {\rm def.}{=}\left({\Lambda^i}_j \right)
\left( {h^j}_\mu \right)_1,\ee where $\left({\Lambda^i}_j
\right)$ is a local Lorentz transformation satisfying \be
\left({\Lambda^i}_j \right)\eta_{ik}
\left({\Lambda^k}_l \right)=\eta_{jl},\ee and
$\left( {h^i}_\mu \right)_1$ is another tetrad field which
reproduces the same metric that is produced by the tetrad field
$\left( {h^i}_\mu \right)$. Now we use Eq.~(22) to rewrite the tetrad
field (12) as:
\be \left( {h^i}_\mu \right)=\left({\Lambda^i}_j \right)
\left({\Lambda^j}_k \right)_1\left( {h^k}_\mu
\right)_d,\ee where \be \left({\Lambda^i}_j \right)= \left(
\matrix{ 1 &0& 0 &0
 \vspace{3mm} \cr  0  & \sin\theta \cos\phi &\cos\theta \cos\phi
 &-\sin\phi \vspace{3mm} \cr
 0  & \sin\theta\sin\phi &\cos\theta\sin\phi &\cos\phi \vspace{3mm} \cr
 0  & \cos\theta & -\sin\theta
 &0 \cr }
\right)\; , \ee \be \left({\Lambda^i}_j \right)_1= \left(
\matrix{\displaystyle\frac{2r^2-2Mr+q^2-\frac{\Lambda r^4}{3}}{2r\sqrt{r^2-2Mr+q^2-\frac{\Lambda r^4}{3}}}
&\displaystyle\frac{2Mr-q^2+\frac{\Lambda r^4}{3}}{2r\sqrt{r^2-2Mr+q^2-\frac{\Lambda r^4}{3}}}& 0
&0
 \vspace{3mm} \cr -\displaystyle\frac{2Mr-q^2+\frac{\Lambda r^4}{3}}{2r\sqrt{r^2-2Mr+q^2-\frac{\Lambda r^4}{3}}}
  & - \displaystyle\frac{2r^2-2Mr+q^2-\frac{\Lambda r^4}{3}}{2r\sqrt{r^2-2Mr+q^2-\frac{\Lambda r^4}{3}}}
  &0 &0 \vspace{3mm} \cr
 0  & 0 &1
 &0 \vspace{3mm} \cr
 0  &0 & 0
 & 1\cr }
\right)\; , \ee and \be \left( {h^i}_\mu \right)_d=\left(
\matrix{\displaystyle\frac{\sqrt{r^2-2Mr+q^2-\frac{\Lambda r^4}{3}}}{r}&
0 &0&0
 \vspace{3mm} \cr 0  &\displaystyle\frac{r}{\sqrt{r^2-2Mr+q^2-\frac{\Lambda r^4}{3}}}
  &0 &0 \vspace{3mm} \cr
 0  & 0 &r
 &0 \vspace{3mm} \cr
0  &0 & 0
 & r\sin\theta  \cr }
\right)\; . \ee The matrices (30) and (31) give local Lorentz
transformations, i.e., they satisfy Eq. (28).  We call matrix (32)
the diagonal form of the tetrad field (12). Finally, equation (31) is the
boost of tetrad field (12).  Matrix (31) shows that the boost
transformation depends on the gravitational mass and  the charge
parameters as well as the cosmological constant. When the charge
parameter is vanishing, the boost transformation depends on the mass
and the cosmological constant and when  charge parameter and
cosmological constant are vanishing then the  boost transformation
depends on the gravitational mass only.

\newsection{Main results and discussion}

 The  $f(T)$ gravitational theory is an amendment   of the teleparallel
 equivalent of general relativity,  which attempts approaching some new observation problems.  To find exact solutions within this theory is not an easy task. We analyzed   the non-vacuum case of    $f(T)$ gravitational theory, with the equations of motion applied to a non-diagonal spherically symmetric tetrad field. Four nonlinear differential equations were obtained.

 To solve these differential equations, we  imposed four constraints (in four unknown functions). The only solution that is compatible with these constraints contains two constants of integration as well as a cosmological constant.   An important property of this solution is that it has a vanishing scalar torsion, i.e. $T=0$, and satisfies the field equations of $f(T)$,  provided that Eq. (25) is satisfied.

 In addition, this solution has an axial-symmetry, i.e., the components of its tetrad field are invariant under the change of the angle $\phi$. We  showed that the two constants of integration related to the gravitational mass and to the charge parameter. The associated metric of this solution gave the Reissner Nordstr\"om-dS spacetime.

To understand the internal properties of the derived solution we rewrote it as two local Lorentz transformations and a diagonal tetrad\footnote{diagonal means the  square  root  of  the  Reissner Nordstr\"om-dS metric.}. It was shown that one of the local Lorentz transformations  is related to  Euler$'$s angle under certain conditions.  This local Lorentz transformation is the same as the one studied in \cite{Ncpl, BMT}. The other local Lorentz transformation depends on the two parameters, the gravitational mass and the charge parameter, as well as the cosmological constant.

To  recapitulate,  any GR solution remains valid in f(T) theories having
 $f_T(0)=1$ whenever the geometry admits a tetrad field with vanishing scalar
 torsion, $T$ \cite{RR11}.
\\
\\

\centerline{\Large{\bf Acknowledgements}}

The author would like to thank anonymous Referees  for their useful
comments that put the paper in a more readable form.

\end{document}